\documentclass[12pt]{article}
\usepackage{scicite}

\usepackage{times}
\usepackage{amsfonts}
\usepackage{xcolor}
\usepackage{siunitx}
\usepackage{graphicx}
\usepackage{nicefrac}
\usepackage{amsmath}
\usepackage{amssymb}

\newenvironment{sciabstract}{%
\begin{quote} \bf}
{\end{quote}}

\newcommand{\kp}{k_{\rm{P}}}

\title{Observation of the diffusive Nambu-Goldstone mode of a non-equilibrium phase transition}

\author
{Ferdinand Claude,$^{1}$ Maxime J Jacquet,$^{1\ast}$ Quentin Glorieux,$^{1}$\\ Michiel Wouters,$^{2}$ Elisabeth Giacobino,$^{1}$ Iacopo Carusotto,$^{3}$\\ Alberto Bramati$^{1\ast}$\\
\\
\normalsize{$^{1}$Laboratoire Kastler Brossel,}
\\ \normalsize{Sorbonne Universit\'{e}, CNRS, ENS-Universit\'{e} PSL, Coll\`{e}ge de France,}\\
\normalsize{Paris 75005, France,}\\
\normalsize{$^{2}$TQC, Universiteit Antwerpen, Universiteitsplein 1, 2610 Antwerpen, Beglium}\\
\normalsize{$^{3}$Pitaevskii BEC Center, INO-CNR and Dipartimento di Fisica, Università di Trento,}\\
\normalsize{via Sommarive 14, I-38123, Trento, Italy}\\
\\
\normalsize{$^\ast$To whom correspondence should be addressed; E-mail: maxime.jacquet@lkb.upmc.fr,}\\ \normalsize{alberto.bramati@lkb.upmc.fr.}
}

\date{}
\begin{document} 
\baselineskip24pt
\maketitle 

\begin{sciabstract}

Second-order phase transitions are governed by spontaneous symmetry breaking, which yield collective excitations with a gapless spectrum called Nambu-Goldstone (NG) modes.
While NG modes in conservative systems are propagating excitations, non-equilibrium phase transitions have been predicted to feature a diffusive NG mode.
We present the first experimental evidence of a diffusive NG mode in a non-equilibrium Bose-Einstein condensate of microcavity polaritons.
The NG mode is observed as a spectral narrowing in the spectroscopic response of the condensate.
Additionally, explicitly breaking  the symmetry causes the opening of a gap in the spectrum and the disappearance of the NG mode.
Our observations confirm the diffusive dynamics of the NG mode of non-equilibrium phase transitions and establish a promising framework to investigate fundamental questions in statistical mechanics.
\end{sciabstract}

From cosmology to particle physics and condensed matter, second-order phase transitions are ubiquitous in classical and quantum statistical mechanics.
They are associated with spontaneous symmetry breaking: beyond the critical point, the symmetric state becomes unstable and multiple degenerate and symmetry-breaking states appear, among which the new equilibrium state is randomly selected~\cite{huang1987statistical,sachdev_quantum_2011}.
When a continuous symmetry is spontaneously broken, as in ferromagnetism and Bose-Einstein condensation for instance, slow twists of the order parameter have a vanishing energy cost.
A gapless and long-lived branch called a Nambu-Goldstone mode then appears in the low-energy part of the collective excitation spectrum ~\cite{nambu_dynamical_1961,goldstone_field_1961,goldstone_broken_1962,lange_nonrelat_1966,TAKAHASHI2015101}.
However, when the symmetry is explicitly broken by an external field, the NG mode is replaced by a gapped branch.

Typically, Bose-Einstein condensation (BEC) is associated with the spontaneous breaking of the $U(1)$ phase symmetry of a bosonic matter field~\cite{Buckingham1968,huang1987statistical,pitaevskii2016bose}.
The excitation spectrum displays a collective NG mode that corresponds to slow twists of the condensate phase (the order parameter).
At thermal equilibrium and in the weakly interacting regime, the collective excitations frequency-wavenumber ($\omega-k$) dispersion has the Bogoliubov form
\begin{equation}
\omega_{bog}(k)=\sqrt{\frac{\hbar k^2}{2m}\left(\frac{\hbar k^2}{2m}+2gn\right)}.
\label{eq:bogoeq}
\end{equation}
Here $m$ and $n$ are the fluid mass and density and $g$ is the interaction constant.
This dispersion satisfies the Goldstone theorem $\omega_{Bog}(k\to 0)=0$ and its low energy and wavenumber behavior $\omega_{bog}(k)=c |k|$ indicates that the NG mode propagates through the condensate as a sound wave with speed $c=\sqrt{\hbar gn/m}$~\cite{pitaevskii2016bose,Steinhauer_spectrum_2002}.

Beyond the specific case of systems at thermal equilibrium \cite{huang1987statistical}, 
symmetry breaking phase transitions generically occur in non-equilibrium systems~\cite{schmittmann_statistical_1995,hidaka_spontaneous_2020}.
For example, non-equilibrium phase transitions underlie coherent-light-emitting devices such as lasers or optical parametric oscillators~\cite{graham_quantum-fluctuations_1968,degiorgio_analogy_1970,graham1970laserlight,grossmann_laser_1971} and non-equilibrium condensation of exciton-polaritons~\cite{kasprzak_boseeinstein_2006,bloch_non-equilibrium_2022}. 
In a polariton condensate at rest, the collective excitation spectrum (measured with respect to the condensate's frequency) was predicted to have the analytical form~\cite{Wouters_nonresonant_NG_2007} 
\begin{equation}
\label{eq:analyticalform}
    \omega_{exc}(k)=-i\frac{\Gamma}{2}+\sqrt{\omega_{bog}(k)^2-\frac{\Gamma^2}{4}},
\end{equation} 
which, at low $k$, reduces to the universal form 
\begin{equation}
\label{eq:universalDiff}
    \omega_{exc}(k)\simeq -i\frac{c^2}{\Gamma}k^2.
\end{equation}
The effective linewidth $\Gamma$ depends on the relative power $P$ above the condensation threshold $P_{th}$ as $\Gamma=\gamma_0(1-P_{th}/P)$, with $\gamma_0$ the bare polariton linewidth.

The NG mode spectrum~\eqref{eq:analyticalform}-\eqref{eq:universalDiff} satisfies the Goldstone theorem and its non-equilibrium characteristics manifest in its real and imaginary parts: the former displays a plateau at zero frequency for a finite range of $k$ around the condensate wavenumber, while the latter grows quadratically with $k$.
Physically, this means that NG excitations do not propagate  across the condensate but monotonically relax with a $k$-dependent rate according to a diffusion equation~\cite{Keeling_nonresonant_NG_2006,Wouters_nonresonant_NG_2007}.
In this work, we demonstrate this general feature of non-equilibrium phase transitions.

\paragraph{The physical system}
We use a semiconductor microcavity in the strong coupling regime, where cavity photons and quantum well excitons form mixed bosonic quasi-particles called polaritons~[Fig.~1~\textbf{A}].
This allows to combine the small effective mass of photons with the sizable binary interactions of excitons, and enables the observation of quantum fluid behaviors~\cite{carusotto_quantum_2013}.

 \begin{figure}[ht]
     \centering
     \includegraphics[width=.9\textwidth]{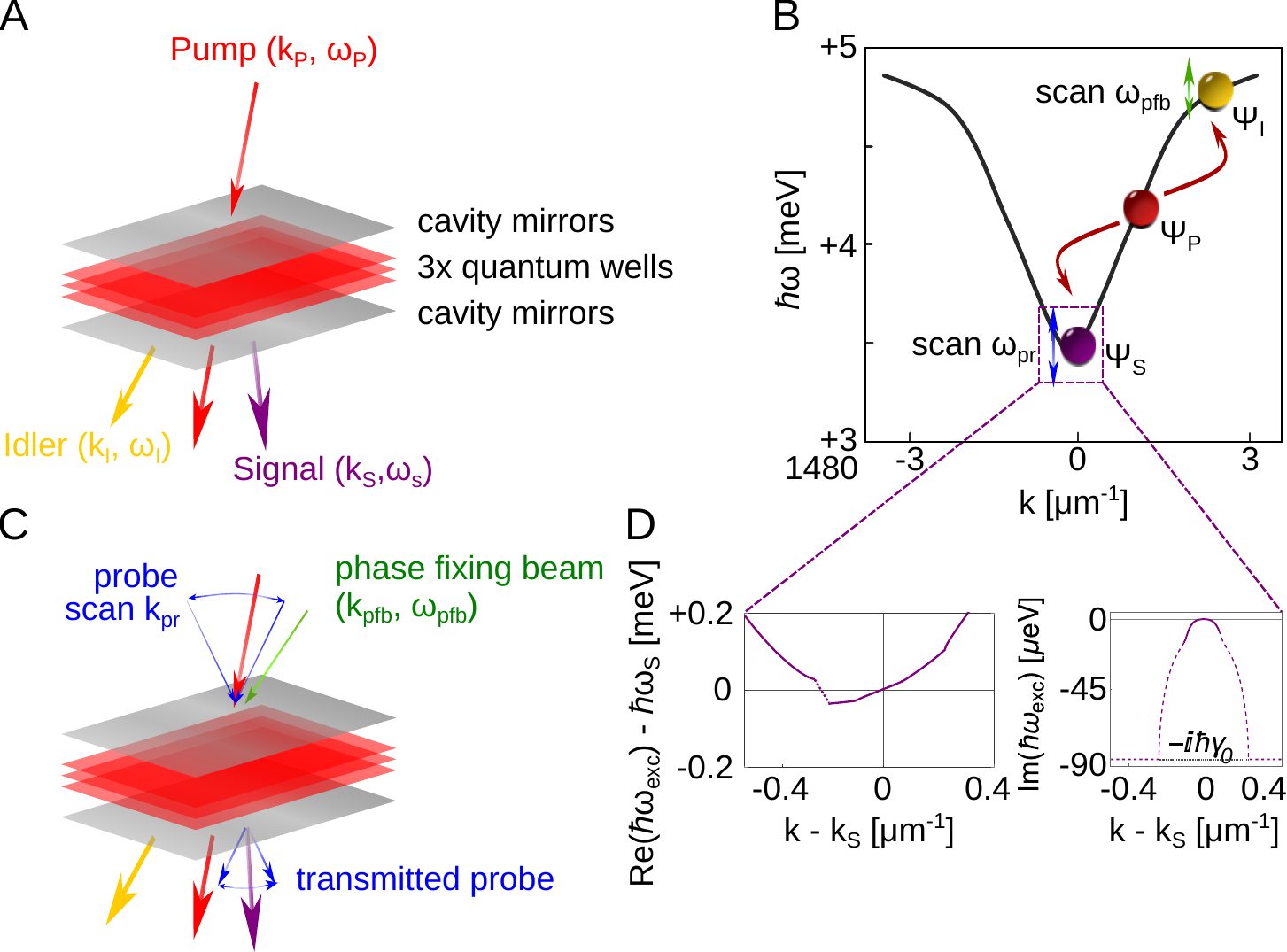}
     \caption{\textbf{Experimental system and theoretical predictions.}
     \textbf{A}, \textbf{B} Parametric oscillation scheme. Pump polaritons $\Psi_P$ (red) at $k_p\approx\SI{1.25}{\per\micro\meter}$ are converted into signal and idler polariton condensates $\Psi_S$ (purple) and $\Psi_I$ (yellow) in the region near $k\approx0$ and at large $k$, respectively.
     \textbf{C} Coherent probe spectroscopy. A probe laser (blue arrow) of tunable angle and frequency ($k_{pr},\omega_{pr}$) scans around ($k_S,\omega_S$). Its transmission yields the real and imaginary parts of the spectrum of elementary excitations.
     A phase fixing beam (green arrow) illuminating the cavity at $k_{pfb}=k_I$ is scanned in frequency $\omega_{pfb}$ around $\omega_I$. When $\omega_{pfb}$ is close to $\omega_I$, the idler's phase is fixed and the U(1) symmetry explicitly broken, leading to the disappearance of the NG mode.
     \textbf{D} Real and imaginary parts of the dispersion of collective excitations of condensate $\Psi_S$ (numerical calculations).
     The NG mode is gapless $\omega_{exc}(k_S)-\omega_S=0$. The quadratic growth of its imaginary part away from $k_S$ indicates a diffusive behavior.
     The real part displays a plateau around $k_S$:  the tilt of the plateau is a consequence of the finite group-velocity $v_S$ of the signal condensate at wavenumber $k_S=\SI{0.11}{\per\micro\meter}$.}
     \label{fig:fig1}
 \end{figure}

Because of the unavoidable radiative and non-radiative losses, a stationary state requires continuous pumping from an external laser source, thus the system has intrinsically non-equilibrium dynamics.
In our experiment, we use parametric interactions to generate the condensate~\cite{baumberg2000parametric,savvidis_angle-resonant_2000}.
As shown in Fig.~\ref{fig:fig1} \textbf{B}, the polariton dispersion features an inflection point around $\SI{1.25}{\per\micro\meter}$.
When a sufficiently strong laser beam pumps the system in the vicinity of this point, stimulated parametric scattering of pump polaritons into a pair of signal/idler modes [Fig.~\ref{fig:fig1} \textbf{B}] gives rise to a parametric oscillation process that leads to the generation of two new polariton fluids $\Psi_{S}$ and $\Psi_{I}$, respectively the signal at $k_S\approx 0$ and $\omega_S$ and the idler at  $k_I=2k_P-k_S$ and $\omega_I=2\omega_P-\omega_S$~\cite{carusotto2005spontaneous}.
Depending on the specific value of $k_P$ around the inflection point, spatial and temporal phase matching lead to the excitation of $\Psi_S$ either at $k_S=0$ or at small yet finite $k_S$. 

This parametric oscillation behavior belongs to the class of non-equilibrium condensation phenomena~\cite{bloch_non-equilibrium_2022}: while the sum of the signal and idler phases is fixed by the nonlinear process to that of the coherently pumped mode at $k_P$, the dynamical equations of the parametric oscillator are invariant under a simultaneous and opposite shift of the signal/idler phases~\cite{graham_quantum-fluctuations_1968,wouters_goldstone_2007}.
Above threshold, the signal/idler phases get locked to a random yet fixed value by spontaneous symmetry breaking.
A NG mode then appears, characterized by a diffusive dispersion at low wavenumber
\begin{equation}
\label{eq:nginpol}
    \omega_{exc}(k)=v_{S}(k-k_{S})-i\frac{c^2}{\Gamma}\left(k-k_S\right)^2\
\end{equation}
that contains a drift term due to the velocity $v_S$ of the condensate, which may be non-zero if the signal condensate is excited at a finite wavenumber $k_S\neq0$.
A numerical calculation of the real and imaginary parts of the dispersion of collective excitations beyond the universal region is shown in Fig.~\ref{fig:fig1}~\textbf{D}.
Physically, the  flat dispersion of the real part at low $k$ corresponds to the tilt of the plateau by the uniform drift at $v_S$, while the quadratic shape of the imaginary part manifests a diffusive behavior.

The consequences of spontaneous symmetry breaking have attracted strong interest since the advent of research in coherent light-emitting devices based on lasing, optical parametric oscillation, or similar processes~\cite{graham_quantum-fluctuations_1968,degiorgio_analogy_1970,grossmann_laser_1971,chow_line_1975,courtois_phase_1991,savvidis_angle-resonant_2000,baumberg2000parametric,baas_quantum_2006}.
Still, the need for spatially large configurations and for high spectroscopic resolutions has so far prevented the experimental observation of the NG mode in the collective excitation spectrum, which is key to completing the field theoretical description of the non-equilibrium phase transition.
Only the temporal signature of critical slowing down in the dynamics of the response to a perturbation has been observed in~\cite{ballarini_observation_2009}.
Meanwhile, if the polariton lifetime is long, the dynamics is approximately conservative: the low $k$ spectrum described by~\eqref{eq:analyticalform} then reduces to the standard Bogoliubov form~\eqref{eq:bogoeq} of equilibrium condensates.
This prevented the observation of the diffusive behavior typical of non-equilibrium systems in~\cite{ballarini_directional_2020}.

\paragraph*{Spectroscopy study of the NG mode}
In order to highlight the non-equilibrium features, we use a sample with a relatively short polariton lifetime consisting of a 2$\lambda$ GaAs/AlGaAs planar microcavity embedding three InGaAs quantum wells at the antinodes of the Fabry-Perot cavity mode.
We measure a Rabi splitting of $\SI{5.07}{\milli\electronvolt}$.
As shown in Fig.\ref{fig:fig1}~\textbf{A}, we pump the cavity at $k_P\approx\SI{1.25}{\per\micro\meter}$, near the inflection point, with a slightly blue-detuned (\SI{834}{\nano\meter}), circularly polarized cw laser of $\SI{100}{\micro\meter}$ waist diameter.
For this value of $k_P$, the signal polaritons appear at a small yet finite wavenumber $k_S=\SI{0.11}{\per\micro\meter}$.

The dispersion of the NG mode is experimentally measured with a recently developed  coherent probe spectroscopy (CPS) method~\cite{claude_high-resolution_2021,claude_spectrum_2023} that provides unprecedented energy and wavenumber resolution.
The CPS is implemented with a second CW coherent laser beam of same polarization as the pump beam and wavenumber $k_{pr}$ in the vicinity of $k_S$.
This configuration allows probing the low-wavenumber collective excitations of the signal polariton condensate $\Psi_S$.
The intensity of the probe beam is set about three orders of magnitude below that of the pump so as to minimally disturb the spontaneous dynamics of the condensate phase (see SM) and is time-modulated at \SI{5}{\mega\hertz} for detection purposes.
The wavevector of the probe $k_{pr}$ is tuned via its incidence angle while its frequency $\omega_{pr}$ is continuously scanned over a range of \SI{100}{\giga\hertz} around the frequency $\omega_S$ of the signal polaritons.
A controllable and $k-$tunable pinhole placed in the reciprocal space of the microcavity selects the transmitted light and directs it on a photodiode.
The electronic signal is then filtered around the modulation frequency with a spectrum analyzer in zero-span mode with a resolution bandwidth of \SI{300}{\hertz}.
In this way, we monitor the probe transmission and identify resonance peaks that appear when $\omega_{pr}\simeq \omega_{exc}(k_{pr})$, corresponding to the collective excitation spectrum.

 \begin{figure}[ht]
     \centering
     \includegraphics[width=.8\textwidth]{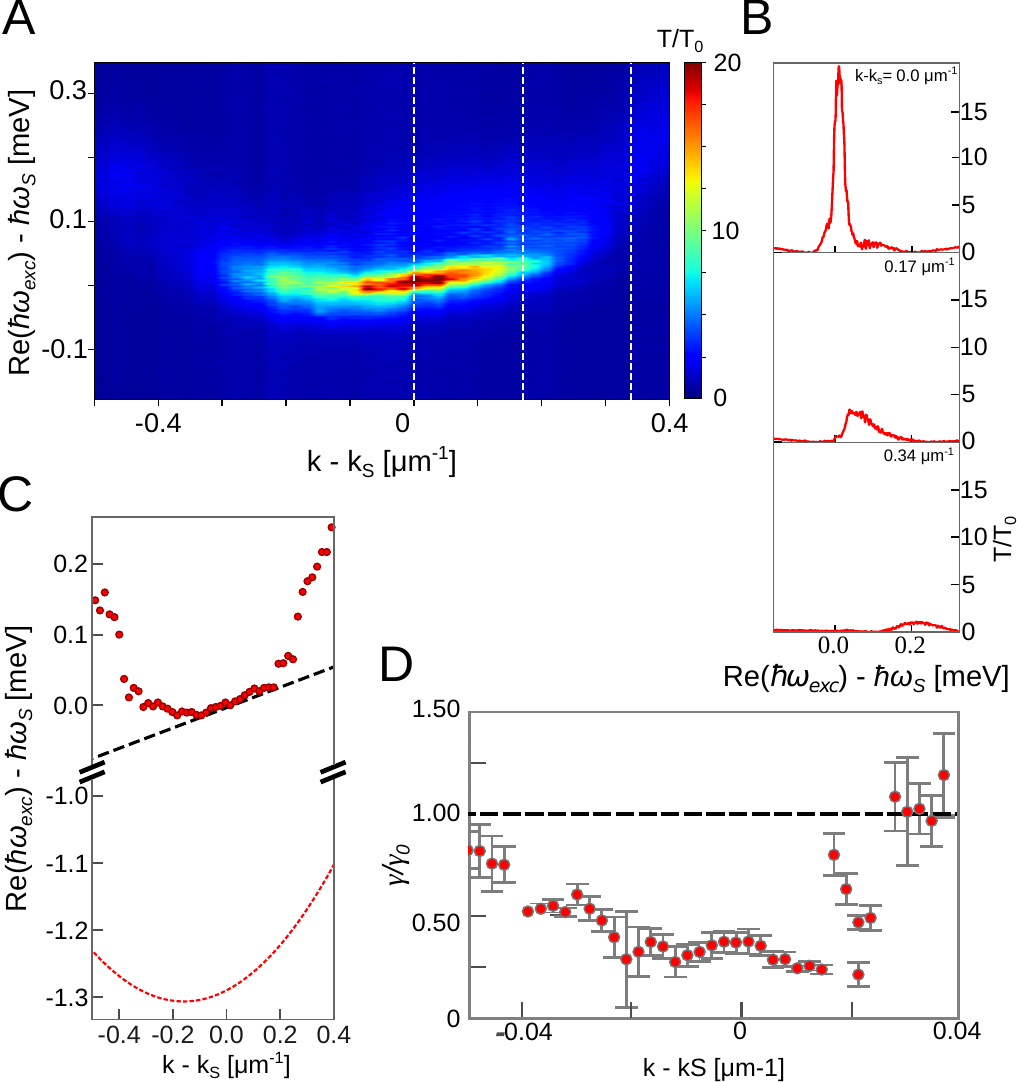}
     \caption{\textbf{Spectrum of collective excitations} for $k_S=\SI{0.11}{\per\micro\meter}$.
     \textbf{A} Color-plot of the normalized probe transmission $T/T_0$.
     \textbf{B} Cuts of the normalized probe transmission taken along the vertical white dashed lines of \textbf{A}.
     \textbf{C} Spectrum of collective excitations (maxima of resonance peaks $\omega_{pr}=\omega_{exc}(k_{pr})$): top, spectrum of excitations in $\Psi_S$; bottom, bare cavity spectrum. The black dashed line highlights the diffusive plateau at $k_S$ (slope $v_S=\SI{0.23}{\micro\meter\per\ps}$).
     \textbf{D} Normalized linewidth $\gamma/\gamma_0$ of the resonance peaks, giving the imaginary part of the Bogoliubov dispersion~\eqref{eq:nginpol}. The error bars correspond to the standard deviation of $\gamma$ (covariance matrix of the Lorentzian fit). For $k-k_S>-\SI{0.45}{\per\micro\meter}$, the fit's squared residuals are above 0.96.
     }
     \label{fig:fig2}
 \end{figure}

A typical transmission spectrum $T$ is shown in Fig.~\ref{fig:fig2}~\textbf{A}; it is normalized to the peak of the bare microcavity transmission $T_0$ measured with the pump beam turned off.
The position and the linewidth of the resonance peak as a function of $k_{pr}$ shown in Fig.~\ref{fig:fig2}~\textbf{C} and \textbf{D} give the real and imaginary parts of the dispersion relation of collective excitations $\omega_{exc}(k)$~\cite{claude_high-resolution_2021}.
The linewidth of the transmission peak changes with $k$: it is approximately equal to the bare cavity linewidth $\gamma_0 = \SI{90}{\micro\electronvolt}$ at large $|k_{pr}-k_S|$, and it decreases until $\gamma/\gamma_0=0.20$ when $k_{pr}$ approaches $k_S$ (the non-zero linewidth is a due to the finite intensity of the probe beam, see SM).
The normalized transmission, $T/T_0$ shows a corresponding trend: it is approximately unity at large $|k_{pr}-k_S|$ and increases up to $T/T_0=18$ when $k_{pr}$ approaches $k_S$.
Indeed, as indicated by the linewidth narrowing, modes with $k-k_S\approx 0$ experience parametric amplification, which manifests as gain in the probe.
The narrow, amplified peak at low $k$ is thus a clear signature of the enhanced response of the NG mode to the probe beam~\cite{wouters_goldstone_2007}.

As seen above, for the chosen value of $k_P$, the signal condensate appears at a finite $k_S$, implying that it flows with a finite group-velocity $v_S$.
From the slope of the diffusive plateau (black dashed curve in Fig.~\ref{fig:fig2}~\textbf{C}), we extract $v_S\approx\SI{0.23}{\micro\meter\per\pico\second}$ .
The finite slope of the plateau proves that the diffusive behavior does not arise from mere elastic scattering of light on cavity disorder or from the finite size of the condensate, both of which would otherwise yield a horizontal plateau.
For the sake of completeness, in the SM we show the dispersion for a different choice of $k_P$ for which the condensate is at rest $v_S\approx 0$ and the plateau is accordingly horizontal, while all other characteristics remain the same.
All these observations provide experimental confirmation to the theoretical prediction of a diffusive NG mode in non-equilibrium systems.

\paragraph*{Explicit symmetry breaking}
Ultimately, the appearance of the NG mode is bound with spontaneous symmetry breaking.
In order to show that this is indeed the physics at play in our experiment, we are now going to explicitly break the symmetry by applying an external field on the polaritons.
In contrast to equilibrium condensates where particle conservation makes it difficult to fix the superfluid phase~\cite{Buckingham1968}, in our optical system a third, phase fixing beam (green beam in Fig.~\ref{fig:fig1}~\textbf{A}) at wavenumber $k_{pfb}$ and frequency $\omega_{pfb}$ close to the idler (yellow beam in Fig.~\ref{fig:fig1} \textbf{A}) can be used to lock the idler phase and, consequently, that of the signal.
This explicitly breaks the signal/idler phase symmetry, which is predicted to result in the disappearance of the NG mode and the opening of a gap in the imaginary part of the spectrum~\cite{wouters_goldstone_2007}.

\begin{figure}[ht]
     \centering
     \includegraphics[width=\textwidth]{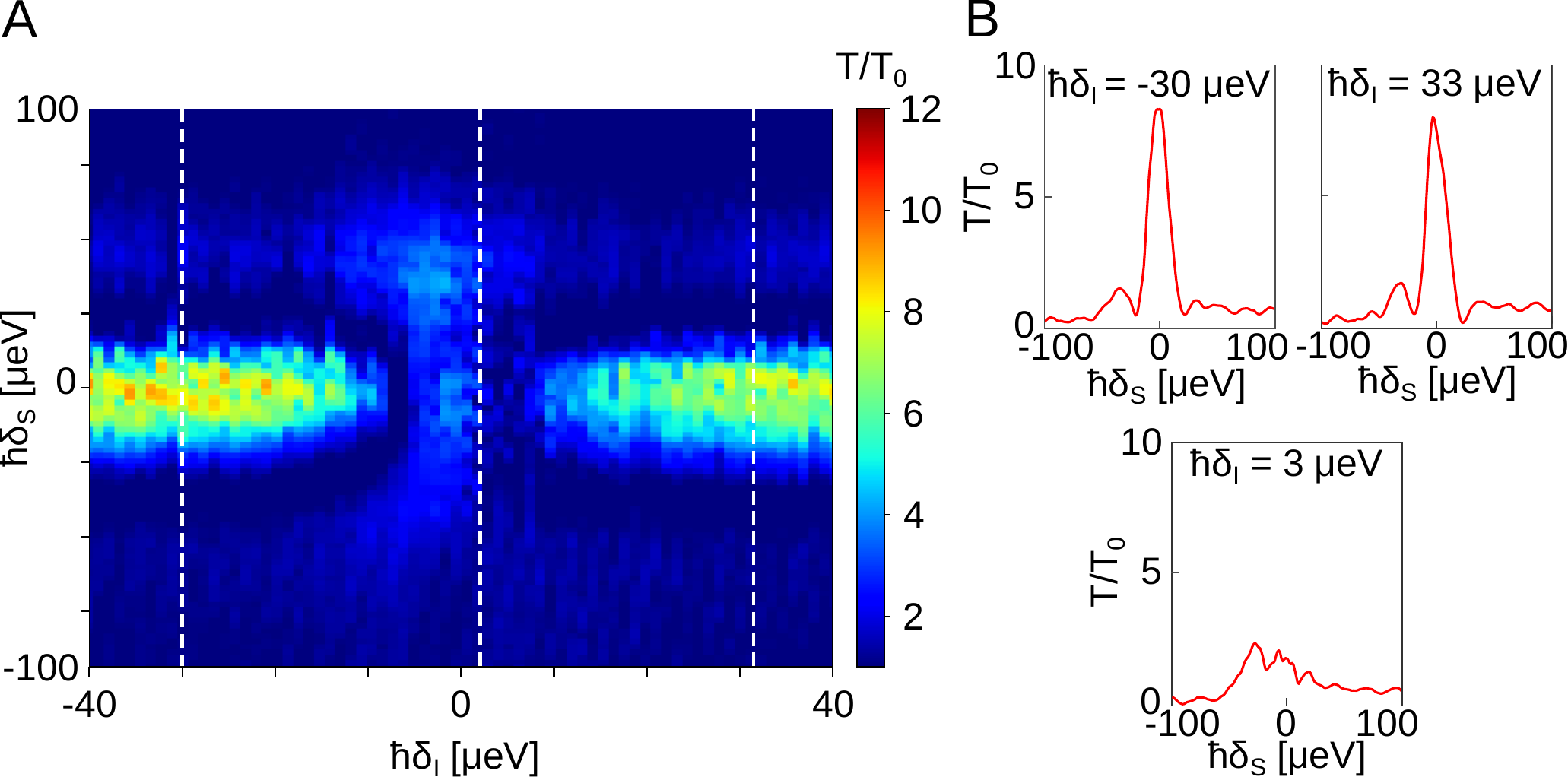}
     \caption{\textbf{Explicit symmetry breaking}. 
     \textbf{A} Color-plot of the normalised probe transmission $T/T_0$ for a probe at $k_{pr}=k_S$ as a function of the frequency detuning between the phase fixing beam and idler polaritons $\delta_I=\omega_I-\omega_{pfb}$ and the frequency detuning between the probe laser and signal polaritons $\delta_S=\omega_S-\omega_{pr}$.
     \textbf{B} Cuts of the normalized probe transmission $T/T_0$ taken along the vertical white lines of \textbf{A}.
     Fixing the idler (and consequently the signal) phase with a third laser on resonance ($\delta_I=0$) results in a broadening of the linewidth and a drop in $T/T_0$.
     This manifests the disappearance of the NG mode under explicit breaking of the phase symmetry.
     \label{fig:fig3}}
 \end{figure}

In our experiment we monitor the probe transmission $T$ in the vicinity of $k_S$ with CPS as the frequency $\omega_{pfb}$ of the phase fixing beam (of wavenumber $k_I$ and moderate intensity, see SM) is scanned around the idler frequency $\omega_I$.
Fig.~\ref{fig:fig3}~ \textbf{A} shows $T$ as a function of $\delta_I=\omega_I-\omega_{pfb}$ and $\delta_S=\omega_S-\omega_{pr}$. 
When the phase fixing beam is detuned in frequency from the idler frequency $\omega_I$, it does not significantly affect the idler field and the spectrum near $k_S,\omega_S$ is characterized by the narrow, amplified peak associated to the NG mode identified in Fig.~\ref{fig:fig2}. 
On the other hand, when the phase fixing beam is on resonance with $\omega_I$, both $\gamma$ and $T$ approximately approach the bare values $\gamma_0$ (corresponding to a gapped imaginary part of the spectrum) and $T_0$, and the NG mode disappears.
This observation unambiguously confirm the origin of the narrow amplified peak in spontaneous symmetry breaking.
Additionally, in the SM we show that the very same phenomenology is observed when the third laser is used to directly fix the phase of the signal polaritons instead of the idler one.

\paragraph*{Conclusion}
In this work, we have experimentally studied the collective dynamics of a non-equilibrium Bose-Einstein condensate of microcavity polaritons.
We have used a new pump-probe spectroscopy method to resolve the intensity and linewidth of the spectrum and observed a narrow, diffusive plateau at low energy.
This is the first experimental evidence of the diffusive nature of the Nambu-Goldstone mode associated with the spontaneous breaking of the phase symmetry at the non-equilibrium condensation phase transition.
As expected, the NG mode disappears when the spontaneous symmetry breaking mechanism is inhibited by fixing the polariton phase with an additional laser beam, which explicitly breaks the symmetry.

Our spectroscopic method will allow to assess corrections to the quadratic scaling of the NG linewidth predicted in Eq.\eqref{eq:universalDiff} for the Kardar-Parisi-Zhang universality class~\cite{ji_temporal_2015} recently observed in the spatio-temporal correlation function of an elongated microcavity polariton condensate in~\cite{fontaine_kardarparisizhang_2022}.
More generally, our results showcase the promise of optical systems and spectroscopic techniques as a versatile platform to investigate fundamental questions in non-equilibrium statistical mechanics~\cite{schmittmann_statistical_1995}, explore non-equilibrium phase transitions that involve several Goldstone branches such as supersolids~\cite{leonard_monitoring_2017,tanzi2019supersolid}, and extend strongly correlated fluids of light~\cite{ma2019dissipatively,clark2020observation} towards new states of matter that have no counterpart in equilibrium systems~\cite{ozawa2019topological}.

\section*{Acknowledgments}
We are thankful to Justin H Wilson for enlightening conversations on NG modes in various field theories. We acknowledge financial support from the H2020-FETFLAG-2018-2020 project ``PhoQuS'' (n.820392).  IC acknowledges financial support from the Provincia Autonoma di Trento, from the Q@TN Initiative, and from the PNRR MUR project PE0000023-NQSTI. MJJ and AB acknowledge financial support from the Sirteq DIM. QG and AB are members of the Institut Universitaire de France. Raw experimental data is available from the corresponding authors upon request.

\section*{Supplementary Materials}

\paragraph*{Excitation scheme}

The experiment is based on the setup described in Ref.~\cite{claude_high-resolution_2021}. It involves the use of three cw laser beams  [Fig.~\ref{fig:fig1} B], with the same circular polarization at the input of the microcavity:
\begin{itemize}
    \item \textbf{the pump laser} -- a Ti:Sapphire laser of frequency and wavenumber adjusted to inject polaritons in the vicinity of the inflection point of the lower polariton branch ($\kp\approx\SI{1.25}{\per\micro\meter}$).
This field  creates the pump mode $\Psi_P$ of the parametric excitation scheme.
Its intensity $I_P = \SI{6.11}{\micro\watt\per\micro\meter\squared}$ is above the parametric oscillation threshold $I_P^{thr} <  \SI{5}{\micro\watt\per\micro\meter\squared} $ at the energy detuning $\omega_P-\omega_{LP}(k_P) = \SI{1.0}{\milli\electronvolt}$ with respect to the polariton resonance at $\kp$ that compensates the blue shift induced by interactions.
In this configuration, optical parametric oscillation leads to the excitation of coherent polariton condensates $\Psi_{S,I}$ at $k_{S,I}$, with $k_S$ small but finite and tunable via $k_P$, see main text.
\item \textbf{the probe laser} -- a Ti:Sapphire laser of frequency and wavenumber set in the vicinity of the signal polaritons ($k_S$).
This field probes the elementary excitations of $\Psi_S$ by scanning their resonance over frequency and wavenumber ranges of $\SI{120}{\giga\hertz}$ ($\SI{0.5}{\milli\electronvolt}$) and  $\pm \SI{1}{\per\micro\meter}$, respectively. 
The probe intensity $I_{pr} = \SI{0.005}{\micro\watt\per\micro\meter\squared}$ is kept three orders of magnitude below the pump intensity so as to minimally disturb the fluid while enabling the observation of the transmission peaks.
The probe is intensity modulated at 5 MHz and detected by a photodiode connected to a spectrum analyzer to filter out the non modulated component due to the unperturbed signal condensate.
\item \textbf{the phase fixing beam} -- a laser diode of frequency $\omega_{pfb}$ and wavenumber $k_{pfb}$ set in the vicinity of the idler condensate ($k_I \approx \SI{2.5}{\per\micro\meter}$).
This field is used to fix the phase of the idler polaritons $\Psi_I$, so as to explicitly break the symmetry of the parametric process.
To this end, its intensity $I_{pfb} = \SI{0.5}{\micro\watt\per\micro\meter\squared}$ is tuned one order of magnitude below the pump and two orders of magnitude above that of the probe.  In one of the next sections of the SM we also show that this same phase fixing beam, tuned to a different wavevector and frequency in the vicinity of $k_S$ and $\omega_S$, can be used to fix the phase of the signal polaritons while monitoring their collective excitation spectrum via the probe beam according to the CPS scheme.
\end{itemize}
The bandwidth of these different lasers is at most \SI{500}{\kilo\hertz} (\SI{2.1}{\nano\electronvolt}), which is several orders of magnitude below the typical linewidth of all modes, and of the observed NG mode in particular (\SI{3}{\giga\hertz}, or \SI{12}{\micro\electronvolt}).
Therefore it does not limit the resolution in the experiments.

\paragraph*{Setting up the parametric oscillation regime}
Here we explain how we image light from the cavity and set up the parametric oscillation regime.

The cavity photoluminescence is imaged in real space to monitor the intensity of emission, while the energy-momentum spectrum is resolved by a spectrometer placed in the Fourier plane.

The pump beam is set in the vicinity of the inflection point of the polariton branch and its intensity increased to reach the threshold of parametric oscillation.
When the system is driven below the parametric threshold, the loss rates are higher than the scattering rate from $\Psi_P$, and several pairs of signal and idler modes are then excited with low amplitude~\cite{drummond_correlations_1990}.
On the contrary, above threshold, a single signal/idler pair dominates, depleting the population of the pump mode at the expense of parametric amplification of all other modes~\cite{wouters_parametric_2007}.

The intensity in all three modes is then maximized by increasing the pump-polariton frequency detuning $\omega_P-\omega_{LP}(k_P)$ towards the blue to compensate the interaction-induced blueshift of the modes.
In the experiment, the optimal value of the detuning is about \SI{1}{\milli\electronvolt} for a beam of waist diameter \SI{100}{\micro\meter} (intensity of \SI{7}{\micro\watt\per\micro\meter\squared}).
The interaction-induced blue-shift of the signal frequency is then of the order of \SI{1.3}{\milli\electronvolt}.

\begin{figure}[ht]
    \centering
    \includegraphics[width=.6\textwidth]{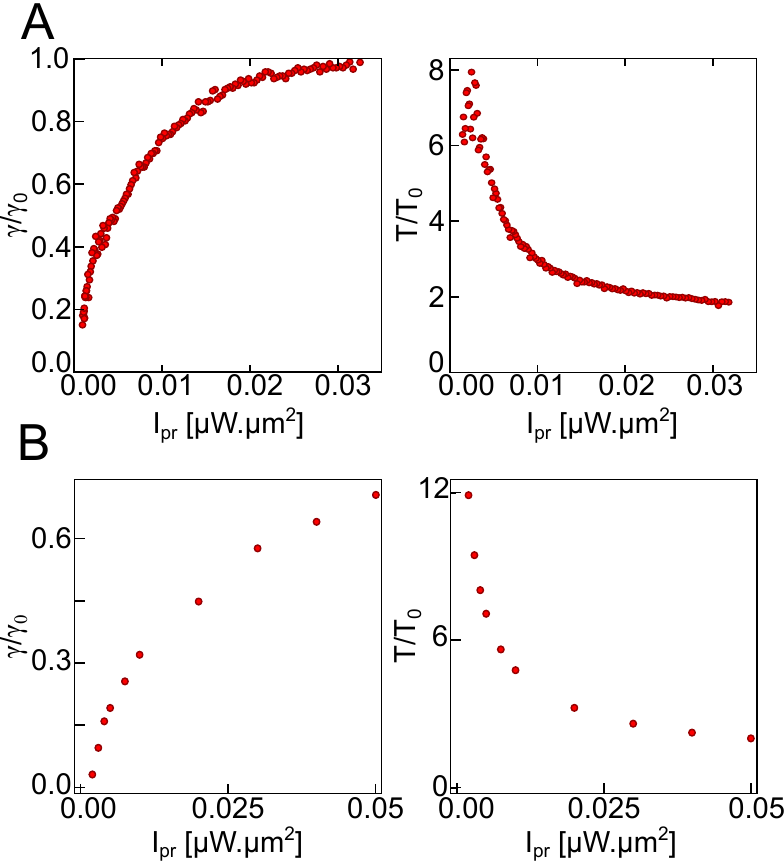}
    \caption{\textbf{Effect of probe intensity on spectral response of the collective excitations in $\Psi_S$}.
    \textbf{A} - \textbf{B} Experimental and numerical data of the normalized peak linewidth $\gamma/\gamma_0$ (left) and transmission $T/T_0$ (right) as a function of the probe intensity in the parametric oscillation regime ($I_P$ = 2.0 $\mu$W/$\mu$m$^2$) at wavenumber $k-k_S=0$. }
    \label{fig:I_probe}
\end{figure}

\paragraph*{Effect of the probe and phase fixing beams on the linewidth of collective excitations}
The probe field creates collective excitations in the signal condensate, but has the side effect of effectively fixing its phase. 
This results in the broadening of the spectral linewidth of the NG mode for increasing probe laser intensity $I_{pr}$, as can be seen in Fig.~\ref{fig:I_probe}. Here, in panel \textbf{A} we show the $I_{pr}$-dependence of the amplitude and full-width at half-maximum (FWHM) linewidth $\gamma$ of the probe transmission signal in the vicinity of $k_S$ where the NG mode is the narrowest.
The perturbation induced by the probe thus effectively limits the minimum observable linewidth, which would theoretically tend towards zero in the limit of vanishing $I_{pr}$, as shown in the experiment (Fig.~\ref{fig:I_probe}~\textbf{B}) and numerical simulations (Fig.~\ref{fig:I_probe}~\textbf{C}).
In the experiments of Figs.~\ref{fig:fig2} and~\ref{fig:fig3} of the main text, we operate with $I_{pr} = \SI{0.005}{\micro\watt\per\micro\meter\squared}$.

A similar balance has to be struck for the phase fixing beam: its intensity is bounded from below by $I_{pr}$ to ensure its dominant role in fixing the idler's phase, while it must be low enough with respect to $I_P$ to avoid disrupting the parametric oscillation process.
In the experiment of Fig.~\ref{fig:fig3}, we operate with $I_I = \SI{0.5}{\micro\watt\per\micro\meter\squared}$.

\paragraph*{Forcing symmetry breaking by pinning the phase of the signal condensate.} 
Spontaneous symmetry breaking in the parametric regime is due to the invariance of the system's Hamiltonian under a simultaneous phase rotation of the signal and idler modes in opposite directions~\cite{wouters_goldstone_2007}.
In the main text, the phase nature of the collective excitations was tested by moving away from this regime: a third beam was used to coherently drive the idler mode, thus fixing its phase and, via the parametric process, that of the signal mode as well.
Parametric oscillation in the idler and signal modes still occurred, but the spectrum of collective excitations on top of the parametrically oscillating state broadened.
Explicit symmetry breaking by setting the phase of the idler and signal mode effectively destroyed the NG mode.
Here we show that the same result may be obtained by directly fixing the phase of the signal mode by bringing the third laser at resonance with it.
\begin{figure}[ht]
    \centering
    \includegraphics[width=.9\textwidth]{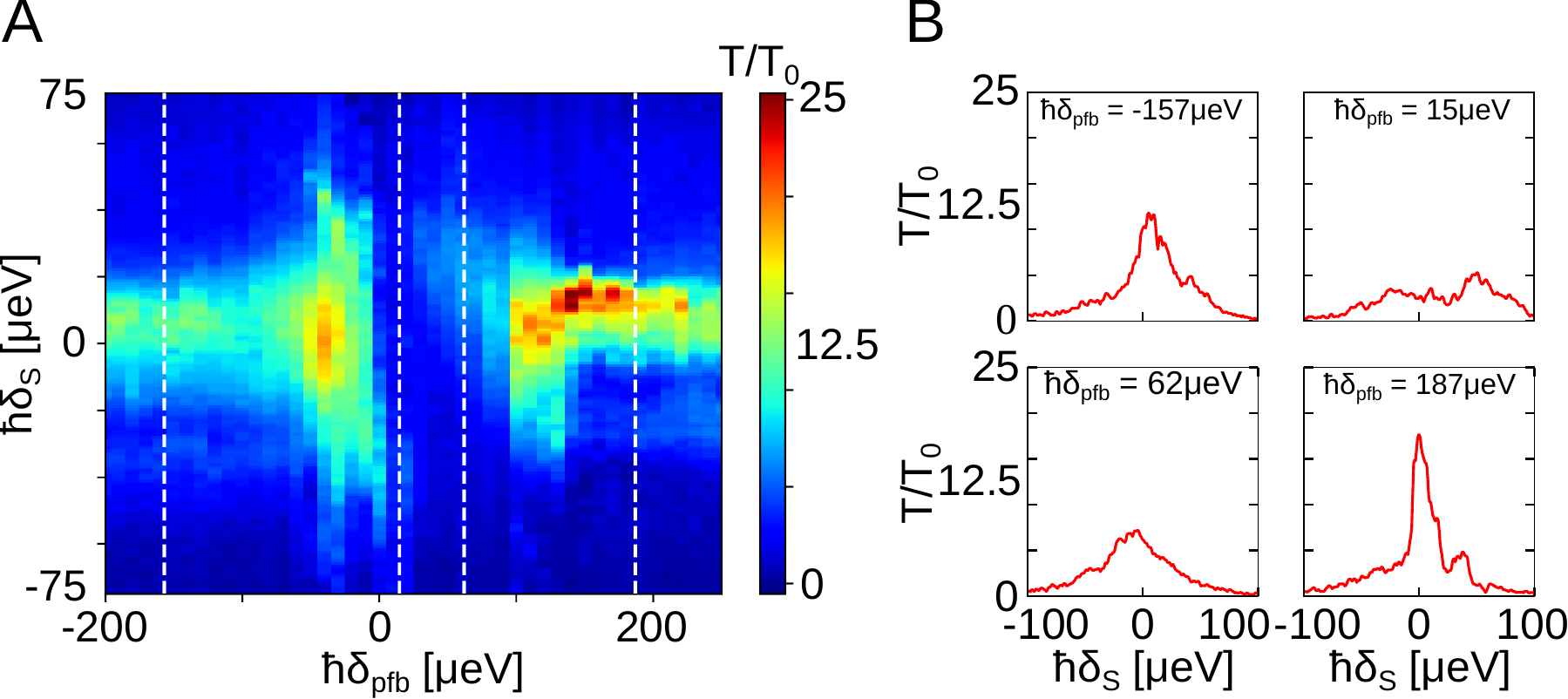}
    \caption{\textbf{Explicit symmetry breaking by fixing the signal's phase}.
     \textbf{A} Color-plot of the normalised probe transmission $T/T_0$ for a probe at $k_{pr}=k_S$ as a function of the frequency detuning between the phase fixing beam and signal polaritons $\delta_{pfb}=\omega_S-\omega_{pfb}$ and the frequency detuning between the probe laser and signal polaritons $\delta_S=\omega_S-\omega_{pr}$.
     \textbf{B} Cuts of the normalized probe transmission $T/T_0$ taken along the vertical white lines of \textbf{A}.
     Fixing the signal phase with a third laser on resonance ($\delta_{pfb}=0$) results in a broadening of the linewidth and a drop in $T/T_0$.
     This manifests the disappearance of the NG mode under explicit breaking of the phase symmetry.
    }
    \label{fig:S_inj_Escan}
\end{figure}

Fig.~\ref{fig:S_inj_Escan}~\textbf{A} shows the transmission $T$ as a function of $\delta_{pfb}=\omega_S-\omega_{pfb}$ and $\delta_S=\omega_S-\omega_{pr}$: For large detunings, $\delta_{pfb}<\SI{0}{\micro\electronvolt}$ and $\delta_{pfb}>\SI{100}{\micro\electronvolt}$, the third laser has no significant effect on the spectral width (see Fig.~\ref{fig:S_inj_Escan}~\textbf{B}) -- the collective excitations are still phase excitations and the NG mode is observed.
On the other hand, at resonance and for a narrow range of  energies $0 <\delta_{pfb} <\SI{100}{\micro\electronvolt}$, the injection of the third laser in the cavity results in spectral broadening up to the bare polariton linewidth $\gamma_0$, which is accompanied by a drop in the transmitted intensity to $T_0$, i.e., a quenching of the probe gain as its parametric amplification is suppressed.
This shows that symmetry breaking may be forced by fixing the phase of the signal mode directly, upon which the NG mode disappears.

\paragraph*{Generating the signal condensate $\Psi_S$ at $k_S=0$}
In the main text, we showed the spectrum of the collective excitations of mode $\Psi_S$ when $k_S=\SI{0.11}{\per\micro\meter}$.
Because of the finite group velocity of $\Psi_S$ at this wavenumber, the spectrum is Doppler shifted according to Eq.~(4) for $v_S>0$ and the diffusive plateau is tilted.
We remark that horizontal diffusive features may result from other phenomena than phase-transition, typically diffusion on a cavity defect. On the other hand, such spurious effects can not be at the origin of a tilted diffusive plateau observed in Fig.~(2), which ensures our interpretation of the observations in terms of NG mode.

\begin{figure}[ht]
\centering\includegraphics[width=.9\textwidth]{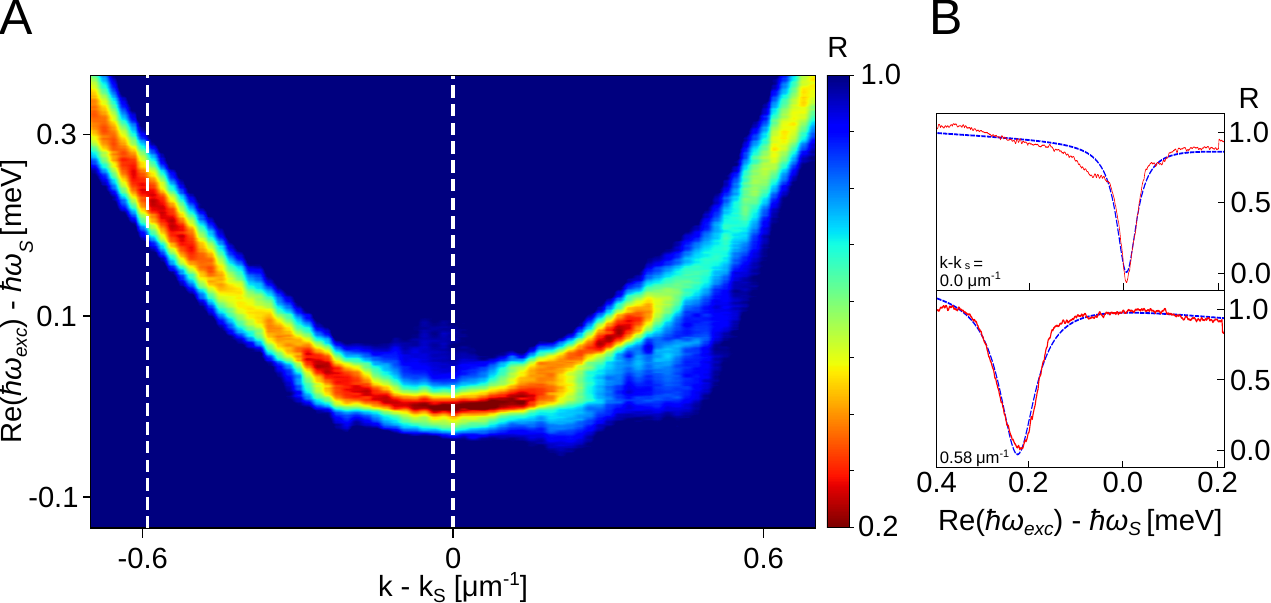}
        \caption{\textbf{Spectrum of elementary excitations near $k-k_S=0$}.
        \textbf{A} Color-plot of the probe reflection $R$.
        \textbf{B} Cuts of the normalized probe reflection taken along the white dashed lines of \textbf{A}, fitted with Lorentzian law plotted in blue dashed lines.
        The detuning and the wavenumber of the pump laser are adjusted to favor the scattering of the polaritons towards a signal mode located as close as possible to $k_S$ = 0.}
    \label{fig:flat}
\end{figure}

For the sake of completeness and as yet another demonstration of the versatility of our methods, here we present experimental data for a different configuration featuring a horizontal diffusive plateau.
We simultaneously varied $k_P$ and the frequency detuning of the pump field and polariton branch so as to operate as close as possible to the inflection point of the polariton branch to generate $\Psi_S$ close to $k_S=0$ such that $v_S\approx0$ and the plateau is horizontal.

Fig.~\ref{fig:flat}~\textbf{A} shows the spectrum of collective excitations in this configuration.
The spectrum is measured from the reflection of the probe in the microcavity.
The spectrum features a horizontal plateau extending from $k=\SI{-0.2}{\per\micro\meter}$ to $k=\SI{0.2}{\per\micro\meter}$.
Fig.~\ref{fig:flat}~\textbf{B} shows that $\gamma\ll\gamma_0$ at low $k$ (down to $\gamma=0.4\gamma_0$), while $\gamma=\gamma_0$ at large $k$.
So the horizontal plateau corresponds to a narrow diffusive mode, which is once again compatible with the excitation of a non-propagative NG mode around $k_S=0$.
Reflection data are bounded between 0 and 1 and cannot show amplification~\cite{claude_spectrum_2023}.

\end{document}